\documentclass[numbers,preprint,5p]{elsarticle}

\makeatletter
\def\ps@pprintTitle{%
  \let\@oddhead\@empty
  \let\@evenhead\@empty
  \let\@oddfoot\@empty
  \let\@evenfoot\@oddfoot
}
\makeatother

\usepackage{tikz}
\usetikzlibrary{shapes,positioning}
\tikzset{node distance = 0.5cm and 0cm}
\usepackage{tcolorbox}
\usepackage{makecell}
\usepackage{booktabs}

\usepackage{url}

\usepackage{breakurl}
\usepackage[breaklinks]{hyperref}

\newcommand{\abbreviations}[1]{%
  \nonumnote{\textit{Abbreviations:\enspace}#1}}

\begin{document}

\begin{frontmatter}

\title{Good Tools are Half the Work:\\
Tool Usage in Deep Learning Projects}

\author[1]{Evangelia Panourgia\fnref{eqcontrib}}

\author[1]{Theodoros Plessas\fnref{eqcontrib}}

\author[1]{Ilias Balampanis}

\author[1,2]{Diomidis Spinellis}

\affiliation[1]{
  organization={Athens University of Economics and Business},
  city={Athens},
  country={Greece}
}
\affiliation[2]{
  organization={Delft University of Technology},
  city={Delft},
  country={The Netherlands}
}

\fntext[eqcontrib]{Both authors contributed equally to this research.}
\abbreviations{DL, deep learning; SE, software engineering; SE4DL, Software Engineering for Deep Learning; MLOps, Machine Learning Operations.}

\begin{abstract}
The rising popularity of deep learning (DL) methods and techniques
has invigorated interest in the topic of SE4DL (Software Engineering for Deep Learning),
the application of software engineering (SE) practices
on deep learning software.
Despite the novel engineering challenges brought on
by the data-driven and non-deterministic paradigm of DL software,
little work has been invested into developing DL-targeted SE tools.
On the other hand, tools tackling non-SE issues specific to DL
are actively used and referred to under the umbrella term
``MLOps (Machine Learning Operations) tools''.\footnote{Our study focuses on deep learning tools.
However, we also examine and reference the wider areas of
machine learning and AI when they are relevant from a
research or practice perspective.}
Nevertheless, the available literature supports the utility of
conventional SE tooling in DL software development.
Building upon previous mining software repositories (MSR) research on tool usage in open-source software works,
we identify conventional and MLOps tools adopted in popular applied DL projects
that use Python as the main programming language.
We furthermore attempt to discern patterns behind the adoption of both types of tooling,
by examining the relative popularity of the SE tools found
compared to their overall popularity,
and by investigating the degree of feature set similarity
in the combinations of MLOps tools we detected.
About 63\% of the GitHub repositories we examined contained at least one conventional SE tool.
Software construction tools are the most widely adopted,
while the opposite applies to management and maintenance tools.
Relatively few MLOps tools were found to be use,
with only 20 tools out of a sample of 74 used in at least one repository.
The majority of them were open-source rather than proprietary.
One of these tools, TensorBoard, was found to be adopted in about half of the repositories in our study.
Consequently, the widespread use of conventional SE tooling
demonstrates its relevance to DL software.
Further research is recommended on the adoption of MLOps tooling,
focusing on the relevance of particular tool types,
the development of required tools,
as well as ways to promote the use of already available tools.
\end{abstract}


\end{frontmatter}

\section{Introduction}\label{sec:intro}
Deep learning techniques have advanced greatly in recent years, leading to the emergence of DL applications in areas such as medicine~\cite{Ker_Wang_Rao_Lim_2017}, finance~\cite{Ozbayoglu_Gudelek_Sezer_2020}, and operations research~\cite{Kraus_Feuerriegel_Oztekin_2020}.

The landscape of software development in DL systems presents novel challenges and opportunities given the \linebreak unique nature of DL application development (i.e. analytical specification of program logic in conventional software vis-a-vis inferring the logic from large datasets), leading to the formation of two distinct related sub-fields of software engineering. The first one is referred to as ``software engineering for deep learning'' (\textbf{SE4DL}), constituting a concept corresponding to the application of SE practices and methods in the development of DL software~\cite{https://doi.org/10.48550/arxiv.2009.08525}. The second relevant concept is referred to as ``deep learning for software engineering'' (DL4SE), in the context of which, existing software development tasks are automated and improved by leveraging DL techniques~\cite{https://doi.org/10.48550/arxiv.2009.08525,KGS22b}. We explicitly treat only the first one in this study.

The shift from traditional software development to deep learning engineering poses unique challenges~\cite{Amershi2019}. In short, this shift derives from the fact deep learning applications are \textbf{data-driven}, whereas traditional software systems are \textbf{logic-driven}~\cite{Zhang2019}. The challenges related to the intersection of software engineering and machine learning have been classified into three categories: \textbf{development}, \textbf{production} and \textbf{organizational} challenges~\cite{Arpteg2018}. We focus on the first category and specifically on the development process of deep learning applications. Development challenges include sub-challenges related to experiment management~\cite{Golovin2017}, maintainability~\cite{Morgenthaler2012}, transparency~\cite{pmlr-v27-bengio12a}, and testing~\cite{Arpteg2018}.          

Other researchers have, in addition to the above, described challenges related to the developers' experience level~\cite[Table II]{Amershi2019}. For example, it was found that the dearth of AI-targeted engineering tools is a major challenge for developers with high experience, whereas access to education and training is the equivalent challenge for developers with low experience~\cite{Amershi2019}.

Our study furthermore relates to MLOps tools, warranting a general introduction to the topic at large. MLOps is considered to be a new field~\cite{Symeonidis2022}, related to both the techniques and the tools being used for the deployment of ML models in production~\cite{Alla2020}. MLOps can, thus, be seen as the convergence of machine learning techniques and DevOps processes and methods~\cite{Symeonidis2022}. DevOps is a group of practices aiming at decreasing the needed time for a software release by means of narrowing the gap between software development and operations~\cite{Fitzgerald2017,Gift2021-bn} following rules and practices such as continuous integration (CI) and continuous delivery (CD)~\cite{Symeonidis2022}.

Despite the great interest which has arisen in recent years regarding DL, leading to a similar effect with respect to SE4DL in both academia and industry, as well as the many empirical studies conducted on conventional software works~\cite{Vidoni2022}, little if any such research has been focused on ML/DL software. We consider the study of tooling used in DL software particularly pertinent to the field, especially in contributing to an improved conceptualization regarding the areas in which DL software differs significantly from conventional software from an engineering perspective. We include MLOps tools to the scope of our study since, as we point out further on, they are not considered strictly distinct from SE4ML/DL ones in the literature. Both categories of tools relate to the automation of engineering tasks relevant to DL software.

Tools are not adopted in a vacuum, leading to a necessity to additionally evaluate which factors might drive their adoption and continued usage. Cataldo et al.~\cite{Cataldo2006} suggest that the adoption of collaboration tools is driven by the complexity of software work management, which increases with the number of developers. This does not necessarily apply only to this specific class of tools; to the extent that engineering tools of all kinds reduce developers' cognitive load by automating ever more tasks, we would expect that SE tools in general are more likely to be adopted both as more participants engage with the software development process, and as the software grows in size. Consequently, we also expect that tool adoption is likewise positively correlated with the number of participants, as well as with the size of the software work itself (measured in e.g. lines of code).

Therefore, our goal for this study is the detection of tools used to aid the engineering of deep learning applications, as well as an investigation of factors related to their adoption. For practical reasons we study open source projects, so as to collect data from a large and diverse set, written in Python specifically due to its overwhelming popularity among machine learning practitioners~\cite{stackoverflow_devsurvey_2022}.

Our research questions are as follows.
\begin{description}
\item[RQ1] How are conventional SE tools used in open-source deep learning applications?
\begin{description}
  \item[RQ1.1] Which SE tools are used in open-source deep learning applications?

  \item[RQ1.2] How does the usage of SE tools in open-source deep learning applications differ from that of Python software at large?
  \end{description}

\item[RQ2] How are MLOps tools used in open-source deep learning applications?
\begin{description}
  \item[RQ2.1] Which MLOps tools are used in open-source deep learning applications?

  \item[RQ2.2] Which combinations of MLOps tools are used in open-source deep learning applications?

  \item[RQ2.3] Which aspects of MLOps tools may affect their adoption?
  \end{description}

\item[RQ3] Are internal pressure factors (project size, number of participants) correlated with a high level of \linebreak SE/MLOps tool adoption in open-source deep learning applications?
\end{description}

We study the aforementioned questions by constructing datasets of public Python DL project repositories on GitHub, and scraping CI/CD workflows, GitHub-reported dependencies, and file contents at large for signs of SE and MLOps tool use. We manually validate the results of the scraping process, and categorize the tools found into subcategories using the SWEBOK knowledge areas~\cite{10.5555/2616205}, as well as the categories reported in the work of Symeonidis et al.~\cite{Symeonidis2022} respectively. We furthermore construct datasets of features for the MLOps tools we found during our study, following the method of the study by Recupito et al.~\cite{10011505}, and use them to explain the adoption of MLOps tooling, isolated or in combinations. Finally, we extract internal pressure metrics and conduct statistical tests to identify any correlations with the number of SE/MLOps tools used.

The contributions of this study comprise
a method for enumerating SE and DL tools in software repositories,
a data set of commonly used tools in DL software, and
novel results regarding the tool-based engineering practices followed in DL projects.

\section{Related work}

The suitability of conventional SE tooling in AI software works is supported by previous SE4ML/DL research. AI-targeted SE tooling, on the other hand, while existing, is still at a nascent stage. Previous mining software repository (MSR) research on SE tool adoption has not focused on DL projects, while also following more restrictive methods relative to this study. Little MSR research has been conducted on MLOps topics.

\subsection{Use of conventional SE tooling in AI software}\label{sec:related_1}

Research on reconciling the differing practices of the SE and ML/DL communities in the development of AI-powered systems through the adaptation of existing processes or the development of novel ones has already been conducted. This line of research often, directly or indirectly, supports the suitability of conventional SE tooling for AI software both at large and in isolation, because the AI component is usually integrated with and supported by a large body of non-AI ones, implementing e.g. stages of an ML pipeline or interfacing with the AI component, a model common in industrial AI software~\cite{10.5555/2969442.2969519}.

Hesenius et al.~\cite{Hesenius2019} present an SE process model for the development of AI software. Recognizing that AI software is usually not deployed as a standalone product~\cite{10.5555/2969442.2969519}, a set of phases to be performed in the scope of and in extension to a conventional SE process model is suggested. The phases involve engineering artefacts (e.g. code, documentation, datasets) as inputs and outputs. We identify an opportunity for the enrolment of SE4ML process artefacts into a software configuration management (SCM) solution.

Serban et al.~\cite{9825909} conducted a mixed-methods empirical study on often-faced challenges and their solutions in re-architecting software systems to support ML components. We study the subset of suggested solutions which explicitly mention practices such as CI/CD, unit and integration testing, infrastructure-as-code, and monitoring as supporting the use of conventional SE tools (i.e. CI/CD platform software, test frameworks and harnesses, observability suites etc.) in ML-powered software.

Focusing on software quality, van Oort et al.~\cite{9474395} conducted an empirical study on the prevalence of code smells in open-source ML software by executing the widely used Python linting tool \texttt{pylint} on GitHub-sourced projects. The findings, which include the pervasiveness of smells such as code duplication and wildcard imports, as well as lacking dependency management practices hurting reproducibility, indicate that using conventional static analysis tools, as well as adopting conventional dependency management tools, can benefit AI software development.

\subsection{AI-targeted SE tools}\label{sec:related_ai}

Notwithstanding commonalities between conventional and AI software engineering, practices specific to engineering AI software inherently suggest a need for SE tooling better suited to AI software development. For instance, according to van Oort et al.~\cite{9474395} and despite the aforementioned possible suitability of static analysis in AI code, some of the rules linting tools implement and enable by default may yet not be relevant to it due to differing practices or lead to outright erroneous generation of warnings due to tool bugs. This leads us to suggest that the proliferation of these tools in AI software development is at least partially dependent on adapting them to the special characteristics such software presents (e.g. by providing AI-targeted sets of rules), along with investing development effort into increasing their reliability for coding practices common in AI software.

AI-targeted SE tools, however, have, at the time of writing not progressed past research efforts. We could only locate two such prototype tools in the area of software quality. One is \texttt{mllint}, developed in the scope of a study on the prevalence of so-called ``project smells'' (a super-set of code smells especially relevant to ML software derived from well-established SE4ML best practices, e.g. use of Data Version Control) in industry ML software by van Oort et al.~\cite{vanOort2022}. The other is MLSmellHound by Kannan et al.~\cite{Kannan2022}, a \texttt{pylint}-based linter that extracts context from the source code (i.e. purpose - ML/non-ML) and uses it to adjust its rule set (by enabling/disabling \texttt{pylint} rules) and output (by adding or removing error messages, rephrasing, and re-prioritizing).

Considerably more effort has been invested into developing AI engineering tools in the area of software testing. The developers of such tools often draw inspiration from and adapt established conventional software testing techniques (e.g. fuzzing~\cite{https://doi.org/10.48550/arxiv.1807.10875, Xie2019, Guo2018}, pairwise testing~\cite{8668044}, concolic testing~\cite{9000035}, mutation testing~\cite{8539073}, white-box testing~\cite{Pei2019}). Such tools still only constitute research prototypes, developed in the scope of measuring the performance of the testing method; through forward snowballing we were unable to find any study concerning their usage in real-world software. We were, moreover, also unable to locate any studies regarding AI-targeted SE tools in other areas (e.g. requirements engineering, design, maintenance).

The examination of grey literature in the field of \linebreak SE4ML/DL, in which we engaged given the lack of comprehensive results in the scholarly literature~\cite{Garousi2019} vis-a-vis tooling, paints a different picture. The popular \texttt{awesome-seml} GitHub repository lists a variety of tools, straddling open-source self-deployed tools and commercial platforms~\cite{githubGitHubSEMLawesomeseml}. These tools aim to support tasks such as data validation, hyperparameter tuning, experiment design and tracking, model packaging, and deployment. These are tasks relevant to the engineering of AI systems which cannot be adequately classified using a conventional SE taxonomy such as the existing~\cite{10.5555/2616205} and upcoming~\cite{SWEBOKv4} SWEBOK knowledge areas.

Tools such as the ones mentioned above are indeed considered in the formal literature to belong to the sub-field of MLOps rather than SE4ML/DL. Symeonidis et al.~\cite{Symeonidis2022}, in their overview of the MLOps subject, present a task-based two-level taxonomy of MLOps tools and use it to classify several instances of them, including many of the ones mentioned in reference~\cite{githubGitHubSEMLawesomeseml}. The authors further identify AutoML tools as being related to MLOps ones and list instances of them.

This overlap in nomenclature might indicate a disconnect between academia and practitioners of SE4ML/DL, yet serves as an impetus for the inclusion of such so-called MLOps tools to the scope of this study. This is our reason for using the sets of tools provided by Symeonidis et al.~\cite{Symeonidis2022} to carry out the second part of our study. We were unable to find other instances of SE4ML/DL being discussed in the grey literature, which is to be expected given the novelty of the subject and the hypothesized overlap with MLOps.

\subsection{Overview of MSR research on tool adoption}\label{sec:related_msr}

Several MSR studies focusing on SE tool adoption in public software repositories have been published in the past. Madeja et al.~\cite{Madeja2021} conducted an empirical study on test case presence and test framework usage in Java repositories on GitHub. The detection of test frameworks was automated using a script, detecting import statements mapped to the frameworks under consideration in source code files.

Kavaler et al.~\cite{Kavaler2019} studied the adoption and replacement of software quality tools and tool chains in JavaScript projects on GitHub using TravisCI as a CI/CD platform, aiming to obtain insights on their impact on metrics associated with the projects (e.g. number of commits, pull requests, issues). The data on tool adoption were gathered from build execution logs rather than workflow specification files, as the authors recognized that such specification files may, in turn, execute secondary ones (e.g. shell scripts, Dockerfiles), implying that a tool might be executed in the scope of a CI/CD job without it necessarily appearing explicitly in the specification file.

Our study differs from the aforementioned ones in several aspects; most importantly, it is the first such study to focus specifically on the presence of tooling in DL software projects.

Our methods to discover instances of SE tool use are not guided by an automated search for strings mapped to tools from a pre-compiled set~\cite{Madeja2021, Kavaler2019}, but rather by an exhaustive automated extraction of data from locations where we consider tool presence to be likely (i.e. CI/CD workflows and project dependencies) combined with manual identification of tool instances in the output data. This reversed search process allows the discovery of tools unique to DL projects (such as the research tools described above), the existence of which we would not necessarily have known beforehand. While our study does not include the extraction of data from secondary locations as in the example conducted by Kavaler et al.~\cite{Kavaler2019}, our alternative investigation of project dependencies balances the scale in terms of comprehensiveness of possible tool instance extraction.

Gonzalez et al.~\cite{Gonzalez2020} conducted a large-scale empirical study on ML repositories on GitHub, considering them as a community and investigating practices relevant to collaborative and independent contribution compared to a control set of conventional software repositories. The authors recognize that applied ML repositories should be considered separately from tool ML repositories; the definition of what is considered to be a ``tool'', however, differs. Whereas D. Gonzalez et al.~\cite{Gonzalez2020} consider tools to include libraries and frameworks, we limit our definition to entities external to application/training code (implying that the adoption of tools might entail writing supporting code, e.g. unit tests or scripts inside CI/CD workflows, but with that code not intended to be compiled into a usable release artefact). In any case, our study focuses explicitly and exclusively on applied DL software works.

We could only locate one other study using GitHub's dependency graph feature as a data source~\cite{10174042}, which we also used extensively both in constructing our initial repository dataset and in the discovery of SE tools; this can be explained in part by the fact that no API to access the data was available at the time of writing, meaning that a specialized scraper would have to be constructed to leverage the feature for large-scale research.

MSR research focusing on the topic of MLOps has also been conducted, albeit to a considerably more limited extent. Calefato et al.~\cite{Calefato2022} studied the adoption of and practices around workflow automation in ML software works on GitHub using GitHub Actions and CML, attempting to locate MLOps-enabled repositories through manual inspection of their respective workflows. Analysis of the project body led to no production-quality ML projects integrating MLOps practices being identified.

While the focus of our study is different to that of the research conducted by  Calefato et al.~\cite{Calefato2022}, the differences in methods present warrant discussion in support of our choices. According to  Calefato et al. the dataset construction section of the paper describes the creation of two separate datasets for each of the CI/CD tools considered. With respect to locating GitHub Actions-enabled ML repositories, whereas Calefato et al. start with a pre-compiled dataset of GitHub Actions-enabled repositories and restrict the number of entries by applying keyword-based filtering, we first collect repositories using a wide variety of DL frameworks and analyze any CI/CD workflows present in them corresponding to a larger number of CI/CD platforms. The comprehensiveness of our study with respect to the detection of MLOps tools is likewise expanded compared to the study conducted by Calefato et al. due to our method, which includes searching the entire contents of each repository for any signs of a large number of MLOps tools. Both of the aforementioned practices adopted by us in this study are supported by the directions for future research provided by Calefato et al.

Regarding our analysis of CI/CD workflows for SE4ML tooling, while Calefato et al. focus on pre-packaged task execution directives (i.e. actions in GitHub Actions), we additionally broaden our study by parsing and analyzing the contents of GitHub Actions \texttt{run} directives, which have the form of CLI shell commands.

\section{Methods}

Our study starts with the construction of a dataset of DL projects guided by popular Python DL frameworks. The next step involves the examination of CI/CD workflows and GitHub-reported dependencies, as well as the carrying out of text searches on the project contents, with the goal of detecting SE and MLOps tools to answer RQ1 and RQ2. The last part considers the extraction of internal pressure metrics from project repositories, namely recent contributors and source lines of code (SLOC) to answer RQ3.

\subsection{Dataset of GitHub deep learning projects}\label{sec:dataset}

\begin{figure}
\begin{center}
\begin{tikzpicture}[font=\small,thick]
\node[draw,
 circle,
 fill] (start) {};
\node[draw,
    rectangle,
    rounded corners,
    minimum width=2.5cm,
    minimum height=0.7cm,
    below=of start] (block1) {Select libraries to be scraped};
\node[draw,
    rectangle,
    rounded corners, 
    minimum width=2.5cm,
    minimum height=0.7cm,
    below=of block1,
] (block2) {Scrape dependent projects of libraries};
\node[draw,
    below=of block2,
    rectangle, 
    rounded corners,
    minimum width=2.5cm,
    minimum height=0.7cm,
] (block3) {Deduplicate projects using multiple libraries};
\node[draw,
    below=of block3,
    rectangle, 
    rounded corners,
    minimum width=2.5cm,
    minimum height=0.7cm,
] (block4) {Remove forks/repositories that are copies of the original};
\node[draw,
    below=of block4,
    rectangle, 
    rounded corners,
    minimum width=2.5cm,
    minimum height=0.7cm,
] (block5) {Filter remaining projects using a threshold of $\leq$ 20 stars};
\node[draw,
    below=of block5,
    rectangle, 
    rounded corners,
    minimum width=2.5cm,
    minimum height=0.7cm,
] (block6) {Filter remaining projects for use of Python};
\node[draw,
    below=of block6,
    rectangle, 
    rounded corners,
    minimum width=2.5cm,
    minimum height=0.7cm,
] (block7) {Filter remaining projects for actively developed ones};
\node[draw,
 circle,
 fill,
 below=of block7] (end) {};
\draw [-latex] 
    (start) edge (block1)
    (block1) edge (block2)
    (block2) edge (block3)
    (block3) edge (block4)
    (block4) edge (block5)
    (block5) edge (block6)
    (block6) edge (block7)
    (block7) edge (end)
;\end{tikzpicture}
\end{center}
\caption{\label{figg}Data collection process.}

\end{figure}
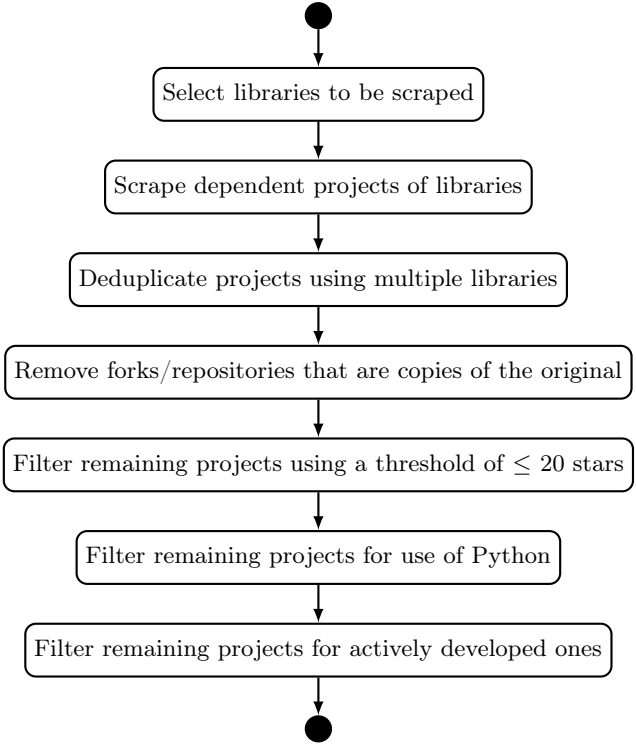

To answer the research questions posited in Section \ref{sec:intro}, we compiled a dataset of DL projects, from which we sampled a subset to examine in depth. The entire process is illustrated in the form of an activity diagram in Figure \ref{figg}.

We set out by locating popular DL libraries and frameworks, which we sourced from academic publications listing such tools~\cite{Erickson2017-kb, Pandey2019-tt, Gheisari2023, Georgiou2022}. We decided to only include libraries/frameworks hosted or mirrored on GitHub in our final list (as opposed to other hosting platforms e.g. GitLab, Codeberg) given its relative popularity, as well as the fact that it was the only one to offer the option of displaying a dependency graph of each repository (listing dependencies/dependents among GitHub repositories), which we utilized in the next step. The results of this step can be found in Table \ref{tab:frameworks}.

\begin{table}[ht]
  \centering
  \caption{\label{tab:frameworks}Popular DL libraries and frameworks.}
  \begin{tabular}{lr}
    \hline
    \toprule
        \textbf{Library/Framework} & \textbf{Source}\\ 
    \midrule
        Blocks	&	\cite{Pandey2019-tt, Gheisari2023}\\
        BigDL &	\cite{Gheisari2023}\\
        Caffe	&	\cite{Erickson2017-kb, Gheisari2023}\\
        Caffe2	&	\cite{Pandey2019-tt}\\
        Chainer	&	\cite{Erickson2017-kb, Gheisari2023}\\
        ConvNetJS	&	\cite{Erickson2017-kb}\\
        CNTK (computational network toolkit) &~\cite{Gheisari2023}\\
        CXXNET	&	\cite{Pandey2019-tt}\\
        Deeplearning4j	&	\cite{Erickson2017-kb, Gheisari2023}\\
        DeepLearnToolBox	&	\cite{Pandey2019-tt}\\
        DeepPy	&	\cite{Pandey2019-tt}\\
        DSSTNE	&	\cite{Erickson2017-kb}\\
        Gensim	&	\cite{Pandey2019-tt}\\
        H2O	&	\cite{Erickson2017-kb}\\
        Hebel	&	\cite{Pandey2019-tt}\\
        Keras	&	\cite{Erickson2017-kb, Gheisari2023}\\
        Lasagne	&	\cite{Pandey2019-tt, Gheisari2023}\\
        MatConvNet	&	\cite{Pandey2019-tt, Gheisari2023}\\
        Microsoft Cognitive Toolkit	&	\cite{Erickson2017-kb}\\
        MXNet	&	\cite{Erickson2017-kb, Gheisari2023}\\
        Neon	&	\cite{Pandey2019-tt}\\
        nolearn	&	\cite{Pandey2019-tt}\\
        Nvidia DIGITS	&	\cite{Erickson2017-kb, Gheisari2023}\\
        PaddlePaddle	&	\cite{Erickson2017-kb}\\
        PyTorch	&	\cite{Erickson2017-kb, Georgiou2022}\\
        SINGA	&	\cite{Pandey2019-tt}\\
        SparkNet &	\cite{Gheisari2023}\\
        Tensorflow	&	\cite{Erickson2017-kb, Georgiou2022, Gheisari2023}\\
        Theano	&	\cite{Pandey2019-tt, Gheisari2023}\\
        Torch	&~\cite{Erickson2017-kb, Gheisari2023}\\
        TFlearn and TFSlim	&~\cite{Gheisari2023}\\
      \midrule
    \end{tabular}
\end{table}

Having identified DL libraries/frameworks, we scraped their dependents (repositories which list the library or framework in question in their manifest files~\cite{github_docs_dependency_graph}) using the listing provided by GitHub in each library's or framework's dependency graph. The data were collected between 2023-10-18 and 2023-10-21, numbering ~366\,000 records along with the \textit{stars} (a user-generated metric of popularity, equivalent to bookmarking) corresponding to each one at the time of collection.

The scraping step was followed by deduplication, during which project repositories with multiple entries in our dataset (due to using more than one of the chosen libraries/frameworks in the first step) were folded into one row, reducing the size of the dataset to ~326\,000 entries. A full left join was then performed between the remaining datasets, and the ``Dataset for GitHub Repository Deduplication''~\cite{spinellis_kotti_deduplication} to exclude any forks and re-uploads of repositories, retaining only the original repositories and further reducing the size to ~324\,000 rows. 

A cutoff of $\geq$20 was applied to the number of stars of each project. This limit has been used in previous MSR research by various authors~\cite{7884605, 10174061, 10.1145/3084226.3084287}. We, similarly to Zhang et al.~\cite{7884605}, use this limit to exclude low-quality projects from our dataset, which we intuitively expect to include personal educational/sample ones. Through this step the size of the dataset was drastically reduced to ~23\,000 projects.

The penultimate step of the process involved, using the relevant endpoint offered by GitHub's REST API~\cite{github_docs_languages_api}, the collection of data on the programming languages used by each project. These data are produced by GitHub's \texttt{linguist} library~\cite{githubAboutRepository}.

Considering that Python is, by far, the most popular programming language among self-declared data scientists and machine learning specialists~\cite{stackoverflow_devsurvey_2022}, we decided to limit the scope of our study to DL projects using Python as their main programming language. In line with our method of choice for collecting programming language data, we achieved that by adapting the very method that the \linebreak \texttt{linguist} library itself employs to report the repository's primary language~\cite{githubLinguistliblinguistrepositoryrb7ca3799b8b5f1acde1dd7a8dfb7ae849d3dfb4cd}; namely, after adding the reported cumulative size of Python and Jupyter notebook files in bytes together, producing a notional ``Python + Jupyter'' language, we associated a primary language with each repository by selecting the one with the largest reported size, and kept only repositories where the associated primary language was our newly produced one.

Given the prevalent practice of integrating AI software components as parts of larger ones~\cite{10.5555/2969442.2969519}, such a reduction of scope serves to limit any cases of false positives in terms of SE tools located, i.e. tools used in the ``non-ML'' part of larger software works (e.g. an image gallery application written mostly in Java, using JUnit, with an object recognition module in Python; the MSR-inspired methods of this study would be inordinately threatened by the prevalence of such cases).

Jupyter notebooks, apart from Python, can be used with a wide variety of other languages through the employment of kernels alternative to IPython~\cite{githubJupyterKernels}. Given, however, that previous research of 2.7 million GitHub-hosted Jupyter notebooks by Källén et al.~\cite{Klln2021} confirmed that Python is, indeed, by far the most commonly used programming language (with fewer than 5\% of the notebooks in the corpus using a different one), we consider our choice to include them, combined with the popularity of Python in the field the projects included in our study belong to, to be reasonable.

Considering the fact that our study focuses on the current state of DL development, we ultimately decided to only include projects that show signs of recent activity, i.e. ones that are being actively worked on. Similarly to the programming language data collection step, we also collected the date of the latest commit on any branch for each repository using GitHub's REST API~\cite{githubRepositoriesGitHub}.
We included the ones that had seen activity in the past two years,
i.e. where the latest commit date was equal or greater to November 1, 2021.
We decided this range based on a similar one found in published research~\cite{10174042}.

Data on programming language use and commit dates were collected on 2023-10-28, with the application of the respective filters reducing the size of the dataset to ~16\,000 records.

Given that our survey focuses on popular applied DL software works, we finally manually reviewed and selected a corresponding subset of repositories to process. We followed one of the approaches used by Gonzalez et al. in~\cite{Gonzalez2020}, wherein scraped repositories were classified as ``Applied'' or ``Tool'' ones by the authors according to the repository description. We deviated from this method in also examining the \texttt{README.md} file of each repository, as we realized that various repositories using terms such as ``toolkit'' or ``framework'' actually provided exclusively model training and inference code, or were collections of models without any independently usable features. We also used the classification ``Other'' for repositories where the reviewing process indicated that their purpose was mostly unrelated to DL (e.g. code for interfacing with robotics hardware). We further applied the label ``Excluded'' for any repositories falling under one of the following categories:

\begin{itemize}
  \item repositories where the majority of the examined artefacts was not in the English language,
  \item repositories that contained educational resources (tutorials, courses) or documentation, and
  \item repositories intended for archival of many, otherwise independent, projects created by the same entity.
\end{itemize}

Two authors independently examined repositories from the dataset in descending GitHub star count order, assigning the aforementioned labels. This process was limited to 10 days due to time considerations, driven by the large size of the dataset, as well as our intention to study popular repositories. During the consolidation process the authors made sure that they had examined the same number of repositories, with the second author examining any missed ones, while any labelling disagreements identified were forwarded to the third author for resolution, implementing a refereeing process. The final number of repositories examined is 700, with the DL repositories included in our study numbering 278.

\subsection{Detection of software engineering tools}\label{sec:SE}

To locate potential instances of use of SE tooling in the selected projects we analyzed:

\begin{itemize}
  \item CI/CD platform workflows, and
  \item GitHub-reported dependencies using its ``dependency graph'' feature.
\end{itemize}

The choice of data items to focus on was motivated by our desire to maximize the probability of detection of conventional SE tools, while not necessarily reducing the presence of potentially unrelated items in the output data (such as software libraries), enabling us to potentially also detect instances of use of unknown to us AI-targeted tools such as those outlined in Section \ref{sec:related_ai}.

With respect to CI/CD platform workflows, we constructed a script to identify the most popular ones in the project body, developing heuristics based on the presence of relevant workflows inside the project files according to the documentation offered by each platform. The platforms whose presence was tested for were sourced from Table I of the paper~\cite{golzadeh_decan_mens_2022} and are reported, along with references to documentation pages we based our work upon, in Table \ref{tab:ci_plat}.

\begin{table}
  \centering
  \caption{\label{tab:ci_plat}Most popular CI platforms.}
  \begin{tabular}{lrl}
    \hline
    \toprule
        \textbf{CI platform} & \textbf{\# Projects} & \textbf{Source}\\ 
    \midrule
        GitHub Actions & 82 &	\cite{githubAboutWorkflows}\\
        Travis & 18 & \cite{travis}\\
        Circle CI & 17 &	\cite{circleciConfigurationIntroduction}\\
        AppVeyor& 2	&	\cite{appveyorBuildConfiguration}\\
        GitLab CI& 1	&	\cite{gitlab}\\
        Azure & -	&	\cite{microsoftCreateYour}\\
        Jenkins & -	&	\cite{jenkinsPipeline}\\
      \midrule
    \end{tabular}
\end{table}
 
By executing the aforementioned script we found that the only CI platforms used in the project body (that is, having at least one project using them) were \textbf{GitHub Actions}, \textbf{TravisCI}, \textbf{CircleCI}, \textbf{AppVeyor}, and \textbf{GitLab CI}.

Having identified the CI platforms of interest, we constructed another script to analyze project CI workflows. Such workflows are usually composed of sequences of shell commands, as well as platform-specific directives expressed in a structured format using a YAML-based DSL. Given that CI workflows automate the execution of tasks relevant to the construction and testing stages of the software development life cycle, these commands and directives may correspond to or contain invocations of and references to diverse SE tools. Our task, therefore, was to extract any such commands and directives possibly referring to SE tools.

We limited ourselves to examining workflows configured to execute on POSIX systems for practical reasons, as the tooling used in our script only supported the \texttt{bash} shell language and ones similar to it, whereas we could not locate similar tooling for the PowerShell or MS-DOS (in batch files) shell languages. Platform-specific directives of interest were found through the examination of CI platform documentation. In cases where a CI platform provides support for environment variable injection in scripting directives (which may, themselves, contain entire commands or parts thereof), the environment variable names were substituted by their values before further processing. We processed specific workflows as follows.

In the case of \textbf{GitHub Actions} we examined the values of \texttt{uses} and \texttt{run} directives. Whereas \texttt{run} blocks contain sequences of shell commands, \texttt{uses} directives execute other workflows as well as packaged bundles of generic CI/CD tasks (the platform's so-called Actions feature). Shell commands under \texttt{run} directives may include references to environment variables listed under the scoped \texttt{env} directives of GitHub Actions workflows.

In the case of \textbf{TravisCI} the directives containing shell commands are limited to the different phases which a CI workflow might execute (such as \texttt{install}, \texttt{deploy}, or \linebreak \texttt{success}/\texttt{failure}, as well as more granular phases identified by the usage of the keywords \texttt{before} and \texttt{after}). Environment variables are located under the workflow-scoped \texttt{env.global} directive.

\textbf{CircleCI} workflows contain commands in \texttt{run} directives, which can either be YAML dictionaries containing \texttt{name} and \texttt{command} keys (of which only the second is of interest), or simple strings. Conditional job steps under \texttt{when} directives were handled as normal ones. Packaged bundles of reusable configurations, similar to GitHub Actions, are imported with the \texttt{orbs} directive, whereas environment variables are used with scoped \texttt{environment} directives.

\textbf{AppVeyor} configurations contain commands under \linebreak phase directives, much like TravisCI, as lists of commands optionally prefixed with the \texttt{sh} directive, and potentially environment variables under the \texttt{environment} directive.

\textbf{GitLab CI} configurations contain commands under \texttt{script} directives, and environment variables under scoped \texttt{variables} directives.

With respect to shell commands, we extracted such sequences and processed them using \emph{bashlex},
a Python port of the \texttt{bash} lexical analyzer~\cite{bashlex}.
Recursively traversing the syntax tree of the block, we first extracted all nodes identified as corresponding to a command (using \texttt{bashlex}'s CommandNode class) and further extracted sub-commands passed to \texttt{docker run} commands as arguments. We then processed the resulting command strings to extract the following pieces of data:

\begin{itemize}
  \item the first word of the node, usually corresponding to a program invocation,
  \item Python packages installed using \texttt{pip},
  \item Python module invocations using the \texttt{python -m} command,
  \item native packages installed using \texttt{apt},
  \item Python packages installed using the \texttt{conda create} command, and
  \item the test harness used in a \texttt{coverage.py} invocation.
\end{itemize}

The above mentioned heuristics were developed by randomly selecting 20 repositories from our dataset and manually examining their CI workflows for traces of specific tool installation to the environment/usage in the building process, and generalizing any located to the rest of the project body.

During the execution of our script, we realized that several workflows made use of \texttt{bash} features unsupported by \texttt{bashlex}, such as the \texttt{[[...]]} conditional construct and arithmetic expressions, or featured non-\texttt{bash} syntax such as the one used for CircleCI pipeline parameters. We resolved this issue by replacing the affected constructs and values with equivalent ones, or with values that would not affect the results of the parsing process, either automatically or manually on a case-by-case basis, keeping a log of all of our manual edits. AppVeyor and GitLab CI workflows were analyzed by manual application of the aforementioned heuristics given that their small number did not call for the extension of our CI workflow parsing script.

GitHub-reported dependencies, sourced using GitHub's dependency graph feature were also scraped and filtered for Python-only dependencies. This was accomplished by consulting GitHub's documentation on the manifest files supported by the feature~\cite{10174042}, identifying manifest file name formats corresponding to Python tooling, and filtering each entry on the name of the manifest file GitHub located it in. Dependencies present only in Poetry lockfiles were also excluded, as they do not constitute direct dependencies of the project being considered (i.e. they are transitive dependencies). The detected dependencies were then merged with the CI workflow-collected data items, at the same time removing any dependencies having a GitHub-reported source repository corresponding to one of the applied DL projects of our study, creating a composite dataset.

The entries of the aforementioned dataset were manually evaluated to confirm their being software engineering tools, as well as to assign them to a SWEBOK knowledge area. Initially this process was aided by consulting the GitHub-reported source repository for entries having one, which proved problematic due to the repository occasionally not being the correct one. We adapted our method by augmenting our dataset with the Python Package Index (PyPI)-reported repository for each entry, and applying the following steps to determine the ``definitive'' source of each entry.

\begin{itemize}
  \item If both the GitHub-reported and the PyPI-reported repositories matched, the definitive source was the reported repository.
  \item If the repositories did not match (including instances of only one of the fields being populated), or no repositories were reported by either source, we determined the definitive source through searching online for the entry and attempting to find a matching website.
\end{itemize}

During the evaluation of the entries in the dataset we also discovered that many entries corresponded to special instances of use of a particular tool (e.g. extensions), or that they constituted evidence of using a specific tool without being the name of tool itself (e.g. GitHub Actions actions). We located those instances, allocated them to the main tool records or defined a more appropriate name for them correspondingly, and revised the project counts of the main records accordingly. Confirmed tools were, finally, allocated to one or more SWEBOK~\cite{10.5555/2616205} categories/knowledge areas. This was achieved through the evaluation of the potential tools by the two first authors, who jointly examined each of them using the aforementioned definitive source. Lastly, the GitHub star count for each non-renamed entry with a presence on GitHub was also scraped (as scraping the star counts of the definitive source of renamed entries would, in virtually all cases, lead to the star count of either an extension of the tool or a GitHub Actions action instead of the star count of the tool itself).

Given that we sourced all of the projects included in this study from GitHub, necessitating their usage of Git as an SCM tool, as well as GitHub as both an SCM tool and project management platform (given its collaboration features such as PRs, and issues), we did not take the last two into account in the scope of this study.

\subsection{Detection of MLOps Tools}\label{sec:mlops}

Our work on detecting instances of usage of MLOps tooling in our project body was based upon the prior work conducted by Symeonidis et al.~\cite{Symeonidis2022}, wherein such tools are enumerated under different categories according to their use case.

\begin{description}
  \item[Data Preprocessing Tools] are divided into \textbf{data labelling} and \textbf{data versioning tools}~\cite{Symeonidis2022}. The first subcategory can be split into more groups related 
to the task being executed. For example, some labelling tools are responsible for 
highlighting particular file types, such as videos and images~\cite{Zhou2017}, while others aid in tagging tasks, such as defining boundary frames and polygonal annotations, and performing semantic segmentation~\cite{Dietterich2002}. Data versioning tools or, alternatively, data extraction tools, manage different versions of data sets and store them in a structured way~\cite{Armbrust2020}. 
  \item[Modelling Tools] support processes related to feature extraction from the raw data set with the purpose of producing optimal data sets for model training. Experiment tracking tools (as a subcategory of modelling tools) are used for monitoring and comparing the different data versions and results derived from each experiment. Hyperparameter tuning is also supported aiming at optimal model \linebreak performance~\cite{pmlr-v28-bardenet13}.
  \item[Operationalization Tools] support the integration of \linebreak ML models in production software~\cite{Savu2011}, the coverage of the entire life cycle of ML applications~\cite{Symeonidis2022}, as well as control of the model. These are important as model performance may decrease after model deployment because of evolving input data~\cite{de_la}.  
\end{description}

We first aggregated the tooling listed into a single \linebreak dataset. For each one of the tools in our dataset we confirmed that it still exists, as well as the expected usage, by conducting a Google search. We discovered that some of the tools could not be found under their original names (due to, for instance, rebranding or acquisition). We decided to include the new names, where applicable, as search keys for those particular tools.

We, additionally, realized that many instances of tool use would probably not be found by looking for the tool's name as-is — many tools are often programmatically interfaced through libraries or other types of software packages. We, therefore, also decided to include PyPI-sourced packages as alternative search keys for each tool, collected by searching for the project's name (as well as any alternative ones, where applicable) on the PyPI website, sorting by relevance, and manually filtering for any that indicated any relevance to the tool in question, either through their name or the short description on the search results page.

The tools listed in reference~\cite{Symeonidis2022} also included platform-level offerings such as Google Cloud Platform (GCP)~\cite{google} and Microsoft Azure~\cite{azure}, each composed of hundreds of domain-specific products. In order to maintain our focus strictly on MLOps-related tools, we decided to exclude these tools from our study.

References to the found tools in our project body were detected by automatically executing a \texttt{grep} command on local snapshots of the project repositories. We configured the command to execute a case-insensitive search, matching exclusively with whole words, while also excluding any binary files.

We furthermore manually confirmed our mining script's results by manually re-executing the command for each of the repositories returned, and confirming that at least one of the matching lines referred to either use of the tool itself or use of a library interfacing with the tool. This process led to the exclusion of projects where the matched name in question was, for instance, found inside a base64 encoded image (e.g. inside Jupyter notebook files), referred to the tool without indicating use by the project in question (e.g. references to tutorials), or was completely unrelated to the tool itself (e.g. general use of the word ``aim'').

Conversely to the manual categorization process followed in subsection \ref{sec:SE} for SE tools, we applied the categories from reference~\cite{Symeonidis2022} to the found instances of MLOps tools. For each tool found in the project body we located its GitHub repository, if existing, and scraped its GitHub stars.

In the scope of RQ2.3, and for each tool found in the project body, we also created feature datasets by consulting their respective documentation web pages, using the 32 principal MLOps tool features identified by Recupito et al.~\cite{10011505}.

\subsection{Extraction of internal pressure metrics}

We extracted a count of recent contributors to the projects examined by cloning their repositories, and executing a script to collect unique contributor names from each repository's Git commits. A cut-off equal to the median number of commits across all projects (146) was used, with us examining only the most recent commits equal to that number, in order to mediate the effect of long-lived larger projects on our data, given that we are studying current practices.

We also extracted the source lines of code (SLOC) metric from each project repository by first converting any Jupyter notebooks in the repositories into Python scripts, keeping only the Python code contained, and executing the \texttt{cloc} utility~\cite{adanial_cloc} on them.

\section{Results and Discussion}

SE tools are used extensively in open-source deep learning projects in Python. A few tools are used in the average repository. Their number confirms our expectations, as well as prior work, related to conventional tool suitability in deep learning. SE tools popular in the Python world are, to a weak extent, likely to also be used for DL software development. MLOps tools are also actively used, though one tool is responsible for most detected instances of tool use. Open-source MLOps tools are preferred. We could not determine any metrics that may drive the adoption of MLOps tools. Statistically significant correlations exist between tool counts and internal pressure metrics.

\subsection{How are conventional SE tools used in open-source deep learning applications?}\label{sec:rq1}

\subsubsection{Which SE tools are used in open-source deep learning applications written in Python?}\label{sec:rq11}

The total number of deep learning projects that was studied was 278. In 102 repositories (\(\sim 37\%\)) no SE tools could be found, with 176 (\(\sim 63\%\)) having adopted at least one SE tool. This finding provides preliminary support for the suitability of SE tooling in DL software works.

\begin{figure}[ht]
\begin{center}
    \includegraphics[width=1\columnwidth,]{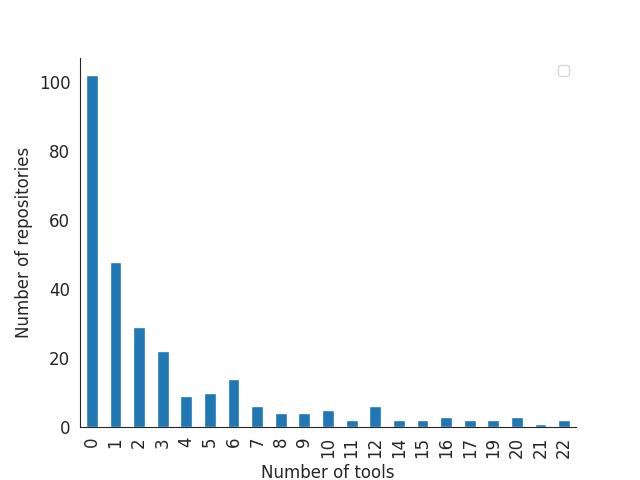}
      \caption{\label{fig:distr}Distribution of number of SE tools in DL projects.}
\end{center}
\end{figure}

With respect to the distribution of the number of SE tools contained in DL projects as depicted in Figure \ref{fig:distr}, most projects appear to use a small number of tools, with 48 (\(\sim 27\%\)) projects using just one tool, and 99 (\(\sim 56\%\)) projects using only 3 tools or less. This may moderate the previous finding, indicating the suitability of only a small number of conventional SE tools to DL software, and would serve to corroborate the dearth of AI-targeted engineering tools constituting a problem for experienced practitioners as reported by Amershi et al.~\cite{Amershi2019}.

\begin{figure}[ht]
\begin{center}
    \includegraphics[width=1\columnwidth,]{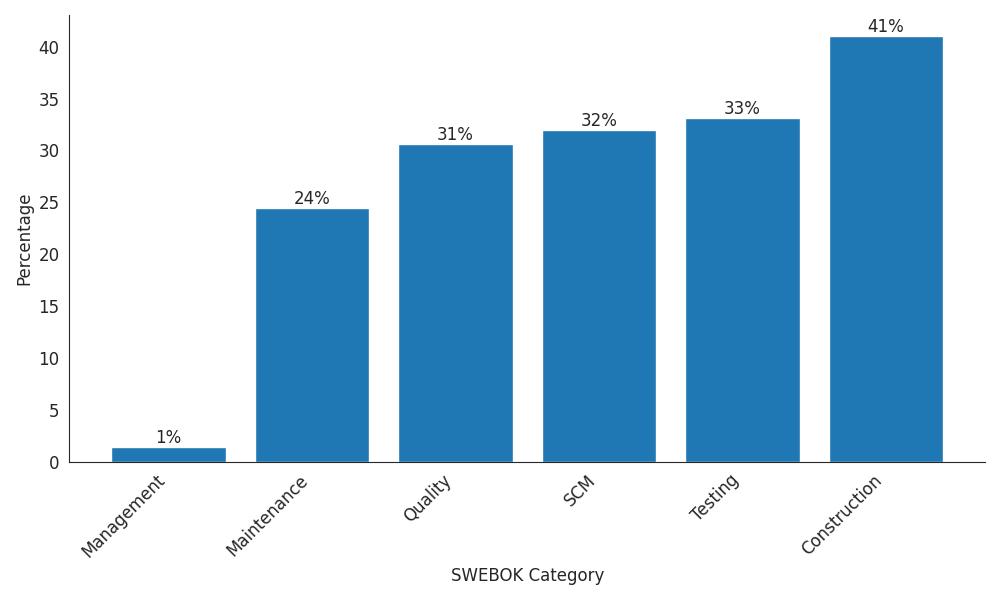}
     \caption{\label{fig:pie_SE}Distribution of used tools by SWEBOK area.}
\end{center}
\end{figure}

Figure \ref{fig:pie_SE} depicts the incidence of tooling in the project body at the SWEBOK category level. Tools belonging to the Construction category can be found in \(\sim 41\%\) of the project body, whereas the Management category is the most underrepresented, with only \~1.5\% of the projects using any such tool; this may be explained by the fact that we have excluded GitHub, which offers a wide variety of software engineering management features, from our study due to its use being a prerequisite for repository inclusion.

No tools belonging to the Software Requirements, Software Design, and Software Engineering Process SWEBOK knowledge areas were detected.

Table \ref{tab:swengtoolinc} lists the six most popular SE tools found in the project body, limited to those with an incidence greater or equal to half of that of the most popular tool. The tools span the spectrum of found SWEBOK categories, with Maintenance being the only missing category.

The presence of \texttt{wheel}, \texttt{setuptools}, \texttt{flake8}, and \linebreak \texttt{sphinx} is not surprising, given that they comprise popular tooling for the packaging of Python projects~\cite{wheel, setuptools}, code style application~\cite{flake8}, and documentation creation~\cite{sphinx} respectively. The presence of \texttt{protobuf} may also be explained by the integration of the data format into Tensorflow~\cite{tensorflowTFRecordTftrainExample}.

The presence of \texttt{pytest}, being the most frequently located SE tool of all, was, however, surprising to us given the limitations of deterministic testing techniques in DL. This may serve to further reinforce the finding in Section \ref{sec:related_ai} on AI testing tools still being at a nascent stage, or may be due to deterministic testing being applied only in parts of the code where it would be suitable (e.g. deterministic data processing code, or other support modules).

\begin{table}
  \centering
  \caption{\label{tab:swengtoolinc}Most popular SE tools detected in GitHub DL projects.}
\begin{tabular}{lrl}
  \toprule
  Tool & Incidence (\%) & SWEBOK category \\
  \midrule
  pytest & 66 (38\%) & Testing \\
  wheel & 40 (23\%) & SCM \\
  flake8 & 38 (22\%) & Quality \\
  protobuf & 36 (21\%) & Construction \\
  sphinx & 34 (19\%) & Construction/Maintenance \\
  setuptools & 33 (19\%) & SCM \\
  \bottomrule
\end{tabular}
\end{table}

While software construction tools such as \texttt{pytest} and \texttt{wheel} show
significant adoption, the under-representation of tools belonging to the Management
and Maintenance categories may indicate a gap in the management of DL projects' life cycles.

There are a number of reasons why tools belonging to the aforementioned categories may be underrepresented.
Continuous experimentation over multiple iterations is a common practice in deep learning initiatives \cite{Amershi2019},
and may be prioritized in place of organized codebase maintenance.
Moreover, the non-deterministic nature of deep learning artefacts may discourage the
implementation of strict quality control procedures. Lack of knowledge or the
perception of complexity in integrating these technologies within a rapidly changing
DL ecosystem could be another factor.

Developers should be encouraged to incorporate complete SE practices throughout the
life cycle of their DL projects in order to close the gap in tool usage. This involves
implementing maintenance tools to aid in iteration and continued support,
and management tools for improved project oversight.
On the other hand, researchers should focus on
creating and developing SE tools that are especially suited for DL settings. This
means improving currently available tools to address the particular difficulties
presented by deep learning, such as reproducible models and data versioning.

\begin{tcolorbox}

\textbf{Answer to RQ1.1} Python DL repositories on GitHub have adopted a variety of conventional SE tools, belonging to the Construction, Testing, SCM, Maintenance, and Quality SWEBOK knowledge areas. Tools belonging to other knowledge areas, however, are not present. More than half of the projects studied use conventional SE tooling, yet the average number of tools in a project is low.

Our findings lend further support to the idea that conventional SE tools are suitable in the development of DL software, but only for specific, targeted use-cases. No AI-targeted SE tools were located.

\end{tcolorbox}

\subsubsection{How does the usage of SE tools in open-source deep learning applications differ from that of Python software at large?}\label{sec:rq12}

We answer this research question  by calculating the correlation of the incidence of the located SE tools with the GitHub stars of the tools, as a proxy of their popularity across Python software at large. We use the Spearman rank correlation coefficient, since we expect both variables to be monotonically increasing.

We express the following null hypothesis:

\textbf{Hypothesis} \textit{There is no correlation between the SE tool incidence in the DL project body and its GitHub star count.}

\begin{table}
  \centering
  \begin{tabular}{@{}lrr@{}}
      \toprule
      & \multicolumn{2}{c}{Hypothesis} \\ \cmidrule{2-3}
      Value & H \\ \midrule
      Correlation Coefficient & 0.30 \\
      P Value& <0.005 \\ \bottomrule
  \end{tabular}
  \caption{\label{tab:hyp12}Spearman correlation results for the hypothesis of RQ1.2.}
\end{table}

As depicted in Table \ref{tab:hyp12}, there exists a weak-to-moderate, statistically significant positive correlation between the variables of the null hypothesis. Therefore, the null hypothesis can be safely rejected, meaning that there is a correlation between a tool's incidence and its star count.

\begin{tcolorbox}

\textbf{Answer to RQ1.2} The presence of SE tooling in Python DL repositories on GitHub is weakly correlated with their overall popularity, indicating that tool adoption may be affected by this factor.
  
\end{tcolorbox}

\subsection{How are MLOps tools used in open-source deep learning applications?}\label{sec:rq2}

\subsubsection{Which MLOps tools are used in open-source deep learning applications?}\label{sec:rq21}

The total number of MLOps tools detected in the selected DL projects was only 20, out of a total of 74. This means that the detected tools in the project body cover just \(\sim 27\%\) of the ones we considered. The number of projects in our study using at least one MLOps tool is 140, corresponding to \(\sim 50\%\) of the project body.

\begin{figure}
\begin{center}
    \includegraphics[width=1\columnwidth,]{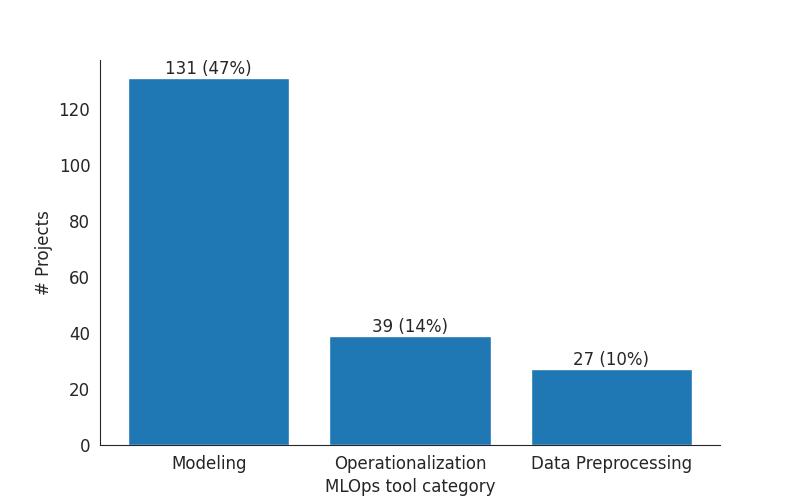}
     \caption{\label{fig:pie_mlops}Distribution of MLOps tool categories.}
\end{center}
\end{figure}

Figure \ref{fig:pie_mlops} depicts the incidence of MLOps tooling in the project body at the category level, i.e. the value of the Category column in Table \ref{tab:mlopst}. By far the most dominant detected category is \textbf{Modelling}, indicating that work falling under it it may be more relevant to open-source DL development. 

The detected MLOps tools are listed in Table \ref{tab:mlopst}. The table's columns are defined as follows:
\begin{itemize}
  \item \textbf{Tool}, the name of the selected MLOps tool,
  \item \textbf{Licensing}, whether the tool is mainly community-driven (i.e. open source) or commercially-driven (i.e. proprietary), with this classification based on reference~\cite{Symeonidis2022}. The value \textbf{OS} represents the term ``open source'', whereas the value \textbf{P} represents the term ``proprietary''. The value \textbf{B}, representing both ``open-source'' and ``proprietary'', is only used in the cases of \texttt{Aim} and \texttt{Weights and Biases}, which can indeed be used as hosted offerings or deployed by the user on their own infrastructure,
  \item \textbf{Incidence (\%)}, the number of DL projects that contain the MLOps tool in absolute and relative terms,
  \item \textbf{FCI} or Feature Completeness Index, the percentage of the 32 features as per Recupito et al.~\cite{10011505}. that the tool offers, 
  \item \textbf{Category}, the first-class MLOps tool categories that the tool belongs to, sourced from reference~\cite{Symeonidis2022}, and
  \item \textbf{Use-case}, the second-class MLOps tool categories that the tool belongs to according to its intended usage, sourced from reference~\cite{Symeonidis2022}.
\end{itemize}

\begin{table*}
  \centering
  \caption{\label{tab:mlopst}MLOps tools detected in GitHub DL projects.}
  \resizebox{\textwidth}{!}{\begin{tabular}{llrrll}
    \hline
    \toprule
        \textbf{Tool} & \textbf{Licensing} & \textbf{Incidence (\%)} & \textbf{FCI} & \textbf{Category} & \textbf{Use case} \\ 
    \midrule
    TensorBoard & OS & 128 (46.0\%) & 40\% & Modelling & Experiment Tracking \\
    OpenVino & OS & 23 (8.3\%) & 21\% & Operationalization & Model Deployment / Serving \\
    Weights and Biases & B & 18 (6.5\%) & 71\% & Data Preprocessing & Data Versioning \\
    Optuna & OS & 13 (4.7\%) & 34\% & Modelling & Hyperparameter Optimisation \\
    mlflow & OS & 10 (3.6\%) & 43\% & Modelling, Operationalization & Experiment Tracking, end-to-end \\
    TorchServe & OS & 8 (2.9\%) & 37\% & Operationalization & Model Deployment / Serving \\
    Comet & P & 7 (2.5\%) & 75\% & Modelling, Data Preprocessing & Data Versioning, Experiment Tracking \\
    Triton Inference Server & OS & 6 (2.2\%) & 34\% & Operationalization & Model Deployment / Serving \\
    Neptune.ai & P & 5 (1.8\%) & 71\% & Modelling & Experiment Tracking \\
    Data Version Control & OS & 4 (1.4\%) & 50\% & Data Preprocessing & Data Versioning \\
    Hyperopt & OS & 4 (1.4\%) & 18\% & Modelling & Hyperparameter Optimization \\
    Amazon SageMaker & P & 4 (1.4\%) & 90\% & Operationalization & end-to-end \\
    TensorFlow Serving & OS & 3 (1.1\%) & 28\% & Operationalization & Model Deployment / Serving \\
    Aim & B & 2 (0.7\%) & 46\% & Modelling & Experiment Tracking \\
    Databricks & P & 2 (0.7\%) & 93\% & Operationalization & end-to-end \\
    Supervisely & P & 1 (0.4\%) & 53\% & Data Preprocessing & Data Preprocessing \\
    SigOpt & OS & 1 (0.4\%) & 50\% & Modelling & Hyperparameter Optimization \\
    Kubeflow & OS & 1 (0.4\%) & 50\% & Operationalization & Model Deployment / Serving \\
    Dolt & OS & 1 (0.4\%) & 34\% & Data Preprocessing & Data Versioning \\
    scikit-optimize & OS & 1 (0.4\%) & 12\% & Modelling & Hyperparameter Optimization \\
    \midrule
    \end{tabular}}
\end{table*}

By far the most dominant MLOps tool is \textbf{TensorBoard}, as it could be verifiably located in \(\sim 46\%\) of the projects of our study. The difference to the second most popular tool, OpenVino, is striking, given that it is only used in \(\sim 8\%\) of the project body. We explain this discrepancy as a result of TensorBoard being one of the two open-source tools in its category (with the second one being Aim) able to be launched straightforwardly from inside a Jupyter environment, while also being the older of the two. On the other hand, tools such as \textbf{Supervisely}, \textbf{SigOpt}, \textbf{Kubeflow}, \textbf{Dolt}, and \textbf{scikit-optimize} are among the least utilized, each appearing in only about 0.4\% of the projects. Their lower incidence ratio can be attributed to the availability of more established alternatives and possibly a narrower range of use cases compared to broader tools, such as those satisfying end-to-end use cases.

Regarding the licensing of the located MLOps tools, 65\% are open-source, whereas 25\% are proprietary. This validates our expectation that open-source DL practitioners are more likely to prefer open-source tooling, even though proprietary tools are, on average more than twice as feature-complete (\(\sim 77\%\)) as open-source offerings (\(\sim 35\%\)). Indeed, of the 14 proprietary tool-using repositories in our study listed in Table \ref{tab:assoc}, 9 (\(\sim 64\%\)) are associated with a for-profit organization.

\begin{table}
  \centering
  \caption{\label{tab:assoc}Proprietary MLOps tool-using repositories and their associated organizations.}
\begin{tabular}{ll}
  \toprule
  Repository & Company \\
  \midrule
  alibaba/AliceMind & Alibaba \\
  ultralytics/ultralytics & Ultralytics \\
  Dao-AILab/flash-attention & - \\
  microsoft/LoRA & Microsoft \\
  ultralytics/yolov5 & Ultralytics \\
  ultralytics/yolov3 & Ultralytics \\
  OpenNMT/OpenNMT-py & - \\
  lllyasviel/ControlNet-v1-1-nightly & - \\
  Picsart-AI-Research/Text2Video-Zero & - \\
  huggingface/diffusers & Hugging Face \\
  awslabs/gluonts & Amazon \\
  databrickslabs/dolly & Databricks \\
  mosaicml/llm-foundry & Databricks \\
  PeterL1n/RobustVideoMatting & - \\
  \bottomrule
\end{tabular}
\end{table}

\begin{figure}[ht]
\begin{center}
    \includegraphics[width=1\columnwidth,]{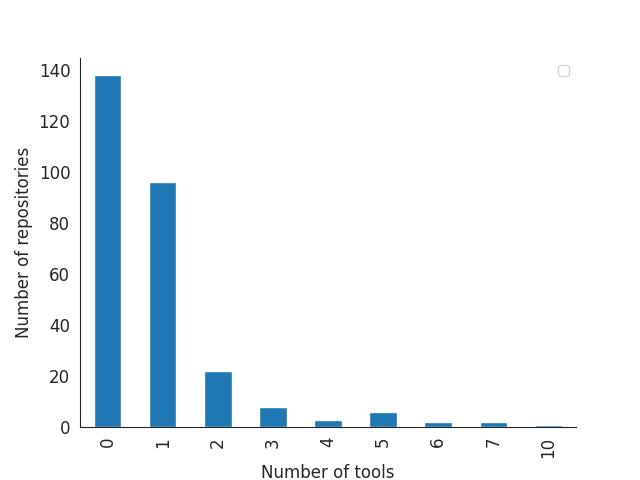}
      \caption{\label{fig:distr_mlops}Distribution of number of MLOps tools in DL projects.}
\end{center}
\end{figure}

Figure \ref{fig:distr_mlops} depicts the distribution of the number of \linebreak MLOps tools contained in DL projects. The distribution is heavily left skewed, with most projects (96, \(\sim 69\%\)) apparently using just one tool (which, again, is most likely to be TensorBoard). This is in line with the similarly heavily skewed preferences of the project body in terms of tooling category, as seen in Figure \ref{fig:pie_mlops}.

\begin{tcolorbox}

\textbf{Answer to RQ2.1} While tools aiding in a wide variety of MLOps tasks are used in many DL repositories, most appear to use Modelling tools, with TensorBoard being by far the prevailing such tool. The majority of projects have adopted just one tool.

The located instances of MLOps tooling use in our study may indicate that such practices have grown more widespread, compared to the findings of previous work~\cite{Calefato2022}.

\end{tcolorbox}

\subsubsection{Which combinations of MLOps tools are used in open-source deep learning applications?}\label{sec:rq22}

We also examined combinatorial tool use, in an effort to elucidate how using different tools together may contribute to facilitating different use-cases under the MLOps life cycle. Given the tool incidence distribution seen in Figure \ref{fig:distr_mlops}, we expected tool combinations to be few in number.

The results validated our expectations, with the combinations of tools found existing in at least two projects, alongside their incidence, their combined feature completeness index, and the categories the tools belong to being listed in Table \ref{tab:combs}. Only four such combinations were found, of which two are composed of only two tools, and with all five featuring Modelling tools and containing TensorBoard, in line with the results of Table \ref{tab:mlopst} and the tool incidence distribution seen in Figure \ref{fig:distr_mlops}.

\begin{table*}
  \centering
  \caption{\label{tab:combs}MLOps tool combinations found in more than one project in project body.}
\begin{tabular}{crrl}
  \toprule
  Tool Combination & Incidence & FCI & Category \\
  \midrule
  \makecell{OpenVino \\ TensorBoard} & 8 & 43\% & Modelling, Operationalization \\
  \midrule
  \makecell{Weights and Biases \\ TensorBoard} & 4 & 71\% & Modelling, Data Preprocessing \\
  \midrule
  \makecell{TorchServe \\ OpenVino \\ TensorBoard} & 4 & 59\% & Modelling, Operationalization \\
  \midrule
  \makecell{Comet \\ Optuna \\ TensorBoard \\ Triton Inference Server \\ OpenVino} & 2 & 81\% & Modelling, Operationalization, Data Preprocessing \\
  \bottomrule
\end{tabular}
\end{table*}

The combination of TorchServe, OpenVino, and TensorBoard contains two tools belonging to the Operationalization category (TorchServe, OpenVino); given the difference in scope between the two tools, this may either indicate that users integrate both of them wishing to support different deployment models (i.e. both the ONNX and OpenVino runtimes), or that they may use different features from each (e.g. TorchServe for deployment, and OpenVino for model compression).

The bottommost combination in the table, composed of six different tools, is used by two company-affiliated repositories, namely \texttt{ultralytics/yolov3} and \linebreak \texttt{ultralytics/yolov5}, while also being the only one to contain a proprietary tool (Comet), providing support to the relevant findings of Section \ref{sec:rq21}.

The average feature completeness index of tool combinations used in two or more projects is \(\sim 64\%\), whereas the same number for isolated tools is \(\sim 48\%\); this, alongside the fact that no tool combinations are composed of tools belonging to only one category, indicates that open-source DL practitioners may indeed adopt different tools for different parts of the MLOps life cycle.

\begin{tcolorbox}

  \textbf{Answer to RQ2.2} MLOps tools are used in combinations, but such instances are few in number. When using tools in such a manner, different tools may be selected to facilitate work across different parts of the MLOps life cycle.
  
\end{tcolorbox}

\subsubsection{Which aspects of MLOps tools may affect their adoption?}\label{sec:rq23}

We answer this research question for both isolated \linebreak MLOps tools, as well as combinations thereof, by examining the correlation between their incidence in our project body and their GitHub stars, as a proxy of their popularity, and the correlation between the incidence and the feature completeness index of the tools. We use the Spearman rank correlation coefficient, since we expect the relationships of the two pairs of variables to be monotonically increasing.

We express the following null hypotheses:

\textbf{Hypothesis 1} \textit{There is no correlation between the isolated MLOps tool incidence in the DL project body and its GitHub star count.}

\textbf{Hypothesis 2} \textit{There is no correlation between the incidence of MLOps tool combinations in the DL project body and their average GitHub star counts.}

\textbf{Hypothesis 3} \textit{There is no correlation between the isolated MLOps tool incidence in the DL project body and its feature completeness index.}

\textbf{Hypothesis 4} \textit{There is no correlation between the incidence of MLOps tool combinations in the DL project body and their combined feature completeness index.}

For practical reasons, given that proprietary tools do not, by definition, publish their source code in GitHub repositories, we are only able to consider open-source tools in testing the first two hypotheses.

\begin{table}
  \centering
  \begin{tabular}{@{}lllll@{}}
      \toprule
      & \multicolumn{4}{c}{Hypothesis} \\ \cmidrule{2-5}
      Value & H1 & H2 & H3 & H4 \\ \midrule
      Correlation Coefficient & 0.12 & -0.32 & 0.07 & -0.95 \\
      P Value & 0.66 & 0.68 & 0.78 & 0.05 \\ \bottomrule
  \end{tabular}
  \caption{\label{tab:hyp23}Spearman correlation results for the hypotheses of RQ2.3.}
\end{table}

Due to the high p-values, as seen in Table \ref{tab:hyp23}, we are unable to reject any of the null hypotheses and must, therefore, accept that there is, indeed, no correlation between MLOps tool incidence, either in isolation or in combinations, and any of the other two metrics of interest.

\begin{tcolorbox}

  \textbf{Answer to RQ2.3} The use of MLOps tooling, in isolation or in combinations, does not appear to be correlated to either the GitHub star count of the tools, or their feature completeness. Given the lack of correlation, it cannot be established that these two metrics affect the adoption of MLOps tooling, meaning that alternative explanations have to be considered.
  
\end{tcolorbox}

\subsection{Are internal pressure factors correlated with a high level of SE/MLOps tool adoption?}

We express the following null hypotheses to provide a statistically sound answer to RQ3:

\textbf{Hypothesis 1} \textit{There is no correlation between recent contributor count and SE tool count.}

\textbf{Hypothesis 2} \textit{There is no correlation between recent contributor count and MLOps tool count.}

\textbf{Hypothesis 3} \textit{There is no correlation between project size (as measured in SLOC) and SE tool count.}

\textbf{Hypothesis 4} \textit{There is no correlation between project size (as measured in SLOC) and MLOps tool count.}

\begin{table}
    \centering
    \resizebox{\columnwidth}{!}{\begin{tabular}{@{}lllll@{}}
        \toprule
        & \multicolumn{4}{c}{Hypothesis} \\ \cmidrule{2-5}
        Value & H1 & H2 & H3 & H4 \\ \midrule
        Correlation Coefficient & 0.39 & 0.20 & 0.52 & 0.50 \\
        P Value& <0.005 & <0.005 & <0.005 & <0.005 \\ \bottomrule
    \end{tabular}}
    \caption{\label{tab:hypotheses}Spearman correlation results for the hypotheses of RQ3.}
\end{table}

All hypotheses were tested using the Spearman correlation coefficient. As shown in Table \ref{tab:hypotheses}, in three of the four cases there exist moderate positive correlations between the two metrics involved, whereas in testing H2, the correlation is weaker but still existent. All correlations are statistically significant given their low p-values. All hypotheses can, therefore, be rejected, meaning that pairwise correlations exist between tools (SE, 9 MLOps) and internal pressure metrics (contributor count, project size).

\begin{tcolorbox}

\textbf{Answer to RQ3.} There exist weak-to-moderate, yet statistically significant, correlations between tool counts and internal pressure metrics (recent contributor count, project size in SLOC) in the project body.

This finding is in line with what was suggested in the examined literature.

\end{tcolorbox}

\section{Threats to validity}

Regarding \textbf{external validity}, the results of our study cannot be generalized to all OSS DL software works - we have only focused on some of the most popular ones, with documentation available in English, developed mainly in Python, sourced exclusively from GitHub.

The results of our study cannot, likewise, be generalized to closed-source DL software works—OSS software is often maintained by enthusiasts, while companies have the resources to invest in upholding SE practices and provisioning MLOps tools and platforms.

Regarding \textbf{internal validity}, the selection of DL \linebreak projects, PyPI packages for MLOps tools, as well as the detection of SE/MLOps tools in the final dataset, were both guided by our personal opinion. As such, we may have excluded projects that would otherwise be considered as DL ones from our study, and conversely included non-DL ones. The same applies for PyPI packages and tools detected. We attempted to address this issue by discussing and resolving divergent opinions among the authors.

With respect to the detection of SE tools, we only examined CI/CD workflows and dependencies reported by GitHub—discounting the influence of personal opinion, \linebreak there could be tools undetected through our methods (e.g. a tool not requiring installation as a dependency, such as UML modelling tooling, tools omitted by accident from the creation of the CI/CD workflow (where available, at that) such as linters, or tools not installed as Python packages). Those tools might have been detected had other artefacts in the repository been analyzed (e.g. configuration files for linters), or had we expanded our method to also include non-Python dependencies from GitHub's dependency graph.

For practical reasons, we could not parse any CI workflows which declared Microsoft Windows as the run-time operating system. Therefore, we may have not located tooling that can only be used under Windows.

The parsing/analysis of the above mentioned CI/CD workflows (e.g. directives to parse in depth) was guided by the documentation provided by their developers. ``Hacky'' instances of platform configuration, using options other than those we found in the documentation, would therefore not lend themselves to parsing (e.g. Jenkinsfiles executing commands by calling a function to e.g. lock a resource).

The tools in our final dataset that were extracted from CI/CD workflows were detected using heuristics we defined—we cannot preclude the possibility of tools existing in the configuration not covered by our heuristics, or of evidence of tooling use that was not detected because of our inability to extract it in a systematic manner (e.g. in the content of \texttt{echo} commands). We attempted to address this issue by manually verifying the obtained results and adjusting correspondingly the employed heuristics.

We only examined the CI/CD workflows themselves, not any other files they could be referring to (e.g. Dockerfiles, shell scripts). Therefore, any tools included in files used in the execution of CI/CD workflow shell commands, for instance, were not detected.

\section{Conclusion and future work}

We examined the usage of conventional SE, as well as MLOps, tooling in public Python DL repositories on GitHub. While the relevance of conventional SE tooling is suggested by the extant literature, no work previous to this study had been done to empirically examine this. We achieved our goal by automating the construction of a dataset of DL repositories, manually selecting the most popular applied DL projects from it, and mining data from sources such as CI/CD workflows, dependency lists, or the repository contents as a whole.

We found that conventional SE tools are indeed used in open-source Python DL projects, with those tools belonging to diverse categories. However, some tool categories were not represented at all. Projects tend to adopt few tools on average, potentially suggesting that their adoption is limited to specific use-cases. DL repositories are somewhat likely to adopt tools that are popular in Python development at large.

We also found that MLOps tools are used widely in the examined repositories, with the tools belonging to that set being adopted by almost half of the repositories. TensorBoard accounted for a considerably large part of the instances detected, being present in about 46\% of the repositories. A clear preference for open-source tooling was detected, while it was also found that repositories adopting proprietary tooling are more likely to be affiliated to a for-profit organization. MLOps tools used in combinations cover different parts of the MLOps life cycle, suggesting the development of MLOps tool-benches. No metrics were found to be correlated with their adoption through statistical testing.

On the other hand, by conducting statistical tests, we detected a correlation between SE/MLOps tool counts and internal pressure metrics (contributor count, project size) in the project body.

Our findings shed light on an understudied area of SE4DL, confirming the suggestions in Subsection \ref{sec:related_1} about the relevance of conventional SE tools in DL software development, as well as providing insights on the adoption of MLOps tools in open-source software.

With respect to conventional SE tools, their apparent relevance in DL software is supported by our findings, lending further credence to the recommendations in reference~\cite{9474395}, namely that the SE tool community should invest efforts in ensuring that such tools produce correct and actionable output for DL software, taking into account its unique characteristics. We further recommend that the aforementioned suggestion be prioritized compared to the further development of AI-targeted SE tools (such as those in Subsection \ref{sec:related_ai}), given that we found conventional ones actively being used in DL software as opposed to AI-targeted ones.

With respect to MLOps tool adoption and development, the prevalence of open-source tools as opposed to commercial ones may either be indicative of their high level of quality or a side effect of their affordability and accessibility. Nevertheless, we believe that vendors of commercial MLOps tools may find it worthwhile to experiment with community outreach programmes, as well as offering free versions of their tools for open-source projects.

The correlation we found between the magnitude of tool use, for both tool categories, and project size, measured both in code and community size, is in line with the reasons that drive tool adoption in conventional software projects \cite{Cataldo2006}. This further provides support to the idea that DL software processes have similar requirements in terms of tooling with conventional ones.

Our suggestions for future research include a qualitative in-depth study of specific OSS DL projects with the goal of determining adherence to SE best practices (e.g. separation of DL pipeline from supporting modules, breaking up DL pipeline into constituent pieces, examining each). We also believe that an MSR study of the use of DL-specific SE tooling~\cite{Giray2021}, with the goal of evaluating whether use of these tools affects various outcomes (e.g. repository popularity, pull request counts), could prove invaluable in further guiding their development.

\section*{Open Science statement}
\label{sec:availability}
The study's replication package and dataset
are available at Zenodo.

\bibliography{sigconf}

\end{document}